\documentclass{aa}  

\usepackage{graphicx}
\usepackage{txfonts}
\usepackage{hyperref}
\usepackage{natbib}
\bibpunct{(}{)}{;}{a}{}{,} 

\begin{document}

   \title{Strong H$\alpha$ emission in the young planetary mass companion 2MASS~J0249$-$0557\,c}


   \author{P. Chinchilla
          \inst{1,2}
          \and
          V. J. S. B\'ejar\inst{1,2}
          \and
          N. Lodieu\inst{1,2}
          \and
          M. R. Zapatero Osorio\inst{3}
          \and
          B. Gauza\inst{4,5}
          }

   \institute{Instituto de Astrof\'isica de Canarias (IAC), Calle V\'ia L\'actea s/n, 38200 La Laguna, Tenerife, Spain.
              \\
              \email{pchinchilla.astro@gmail.com}
         \and
             Departamento de Astrof\'isica, Universidad de La Laguna (ULL), 38205 La Laguna, Tenerife, Spain.
             \and
             Centro de Astrobiolog\'ia (CSIC-INTA), Ctra. de Ajalvir km 4, 28850 Torrej\'on de Ardoz, Madrid, Spain.
             \and
             Centre for Astrophysics Research, University of Hertfordshire, College Lane, Hatfield AL10 9AB, UK.
             \and
             Janusz Gil Institute of Astronomy, University of Zielona G\'ora, Lubuska 2, 65-265 Zielona G\'ora, Poland.
             }

   \date{Received 24 June 2020; accepted 10 November 2020}

 
  \abstract
  {}  
   {Our objective is the optical and near-infrared spectroscopic characterisation of 2MASS J0249$-$0557 c, a recently discovered young planetary mass companion to the $\beta$ Pictoris ($\sim$25 Myr) member 2MASS J0249$-$0557.}
   {Using the Visible and Infrared Survey Telescope for Astronomy (VISTA) Hemisphere Survey (VHS) and the Two Micron All Sky Survey (2MASS) data, we independently identified the companion 2MASS~J0249$-$0557\,c. We also obtained low-resolution optical spectroscopy of this object using the Optical System for Imaging and low-intermediate-Resolution Integrated Spectroscopy (OSIRIS) spectrograph at the Gran Telescopio Canarias (GTC), and near-infrared spectroscopy using the Son of Isaac (SofI) spectrograph on the New Technology Telescope (NTT).}
   {We classified 2MASS~J0249$-$0557\,c with a spectral type of L2.5$\pm$0.5 in the optical and L3$\pm$1 in the near-infrared. We identified several spectroscopic indicators of youth both in the optical and in the near-infrared that are compatible with the age of the $\beta$ Pictoris moving group: strong absorption due to oxides, weak alkaline atomic lines, and a triangular shape of the $H$-band pseudo-continuum. We also detect a strong H$\alpha$ emission, with a pseudo-equivalent width (pEW) of $-90^{+20}_{-40}\AA$, which seems persistent at timescales from several days to a few years. This indicates strong chromospheric activity or disk accretion. Although many M-type brown dwarfs have strong H$\alpha$ emission, this target is one of the very few L-type planetary mass objects in which this strong H$\alpha$ emission has been detected. Lithium absorption at 6708~$\AA$ is observed with pEW $\lesssim$ 5~$\AA$. We also computed the binding energy of 2MASS~J0249$-$0557\,c and obtained an (absolute) upper limit of $U=(-8.8\pm4.4) 10^{32}$~J.}
   {Similarly to other young brown dwarfs and isolated planetary mass objects, strong H$\alpha$ emission due to accretion or chromospheric activity is also present in young planetary mass companions at ages of some dozen million years. We also found that 2MASS~J0249$-$0557\,c is one of the wide substellar companions with the lowest binding energy known to date.}

   \keywords{brown dwarfs -- planetary systems -- binaries:visual -- open clusters and associations: individual: $\beta$ Pictoris   -- Stars: pre-main sequence
               }
\titlerunning{Strong H$\alpha$ in the young planetary mass companion 2M0249-0557 c}
   \maketitle
%

\section{Introduction}

One of the main challenges in astronomy today is understanding the formation and evolution of planetary systems. The characterisation of planets of different ages is key to providing a more accurate picture of the different stages of their formation. However, this characterisation is very difficult because planets are frequently too close to their parent stars and are too faint compared to them to be directly imaged. One possible solution for this issue is finding wide ($>$50--100\,AU) planetary mass companions, which can serve as analogues to the close-orbit giant planets. Because these objects have a wide separation from their primary star, they can be fully characterised using spectroscopy \citep[e.g. ][]{Aller2013, Bowler2014, Deacon2014, Gauza2015, Dupuy2018, Chinchilla2020}.

$\beta$ Pictoris is a young moving group (YMG) named after its brightest and most famous member, the A-type star $\beta$ Pictoris. The group was first discovered by \citet{Zuckerman2001}, who identified 17 young stellar systems moving through space together with the main representative of the group. With an age of $\sim$20--25 Myr \citep{Barrado1999, Mentuch2008, Binks2014, Malo2014, Mamajek2014, Bell2015, Shkolnik2017} and a mean heliocentric distance of $\sim$40 pc \citep{MiretRoig2018}, its youth and proximity make the $\beta$ Pictoris YMG one of the best laboratories for the search and characterisation of young substellar companions.

One of the most recently discovered wide planetary mass companions is 2MASS~J02495436$-$0558015 \citep[hereafter 2MASS~J0249$-$0557\,c; ][]{Dupuy2018}. This is a $11.6_{-1.0}^{+1.3}\,M_{\rm Jup}$ object orbiting at a separation of $\sim$1950 AU from the M6 member of $\beta$ Pictoris YMG 2MASS~J02495639$-$0557352. The primary itself has been resolved by the Keck Laser Guide Star (LGS) adaptive optics into a close (2.17$\pm$0.22\,AU, 0.04\arcsec) brown dwarf pair (hereafter 2MASS~J0249$-$0557 AB), with estimated masses of $48_{-12.0}^{+13}\,M_{\rm Jup}$ and $44_{-11.0}^{+14}\,M_{\rm Jup}$, respectively \citep{Dupuy2018}.

We present the independent discovery and new optical and near-infrared characterisation of 2MASS~J0249$-$0557\,c. Section 2 describes the main characteristics of the system. In Section 3 we present the independent identification of this object.  Section 4 describes the optical and near-infrared spectroscopic observations obtained for the planetary mass companion, and section 5 presents the main results. Finally, Section 6 provides the summary and final remarks.

\section{2MASS J0249$-$0557 system}

The primary, 2MASS~J0249$-$0557, was included in \citet{Shkolnik2017} as a new confirmed member of $\beta$ Pic moving group. They determined a spectral type of M6 and found very low gravity signs in its near-infrared spectrum. They measured a radial velocity of 14.42 $\pm$ 0.44 km s$^{-1}$, a  lithium pseudo-equivalent width (pEW) of 0.59$\pm$0.05~$\AA$ and a strong H$\alpha$ emission line with a pEW of $-$11.6$\pm$0.1~$\AA$. 

\citet{Dupuy2018} recently published the discovery of 2MASS~J0249$-$0557\,c as a wide (39.959$\pm$0.005\arcsec, 1950$\pm$200~AU) companion to 2MASS~J0249$-$0557, one of the targets from the Hawaii Infrared Parallax Program. They also found that the primary is a tight brown dwarf binary, revealing that 2MASS~J0249$-$0557ABc is a bound triple system. Using the Keck/NIRC2 instrument and the laser guide star adaptive optic system (LGS AO), they resolved the primary into two similar magnitude objects located at an angular separation of 44.4$\pm$2\,mas (equivalent to 2.17$\pm$0.22\,AU). They determined a near-infrared spectral type of L2 VL-G for the wide companion and masses of 48$^{+13}_{-12}$ and 44$^{+14}_{-11}$\,M$_{\mathrm{Jup}}$ for the tight binary and 11.6$^{+1.3}_{-1.0}$\,M$_{\mathrm{Jup}}$ for the wide companion. They measured the parallax and proper motions, and confirmed the membership of this system in the $\beta$\,Pic moving group. They showed that the companion 2MASS~J0249$-$0557\,c is physically bound, which means that this is one of the few triple substellar systems known to date.

  \begin{figure}
   \centering
   \includegraphics[width=6cm]{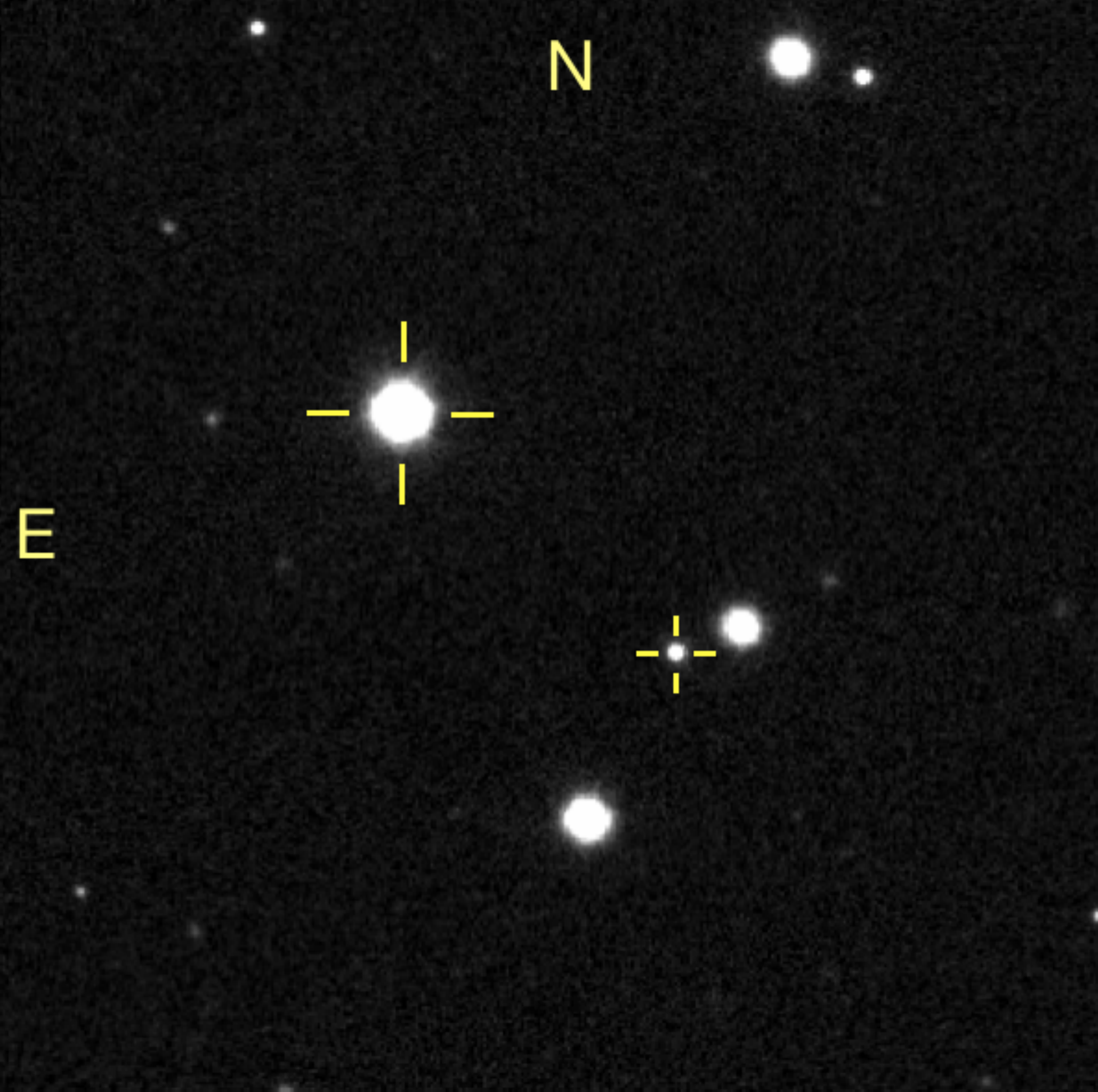}
   
   \caption{\textit{J}-band VHS finding chart for the 2MASS J0249$-$0557 system. The system components are marked with crosses. The field of view is 2\arcmin $\times$ 2\arcmin and the orientation is north up and east to the left. The angular separation between the components is 40.0\arcsec.}
              \label{fig:finding_chart}%
    \end{figure}

 \begin{figure}
   \centering
   
   \includegraphics[width=7cm]{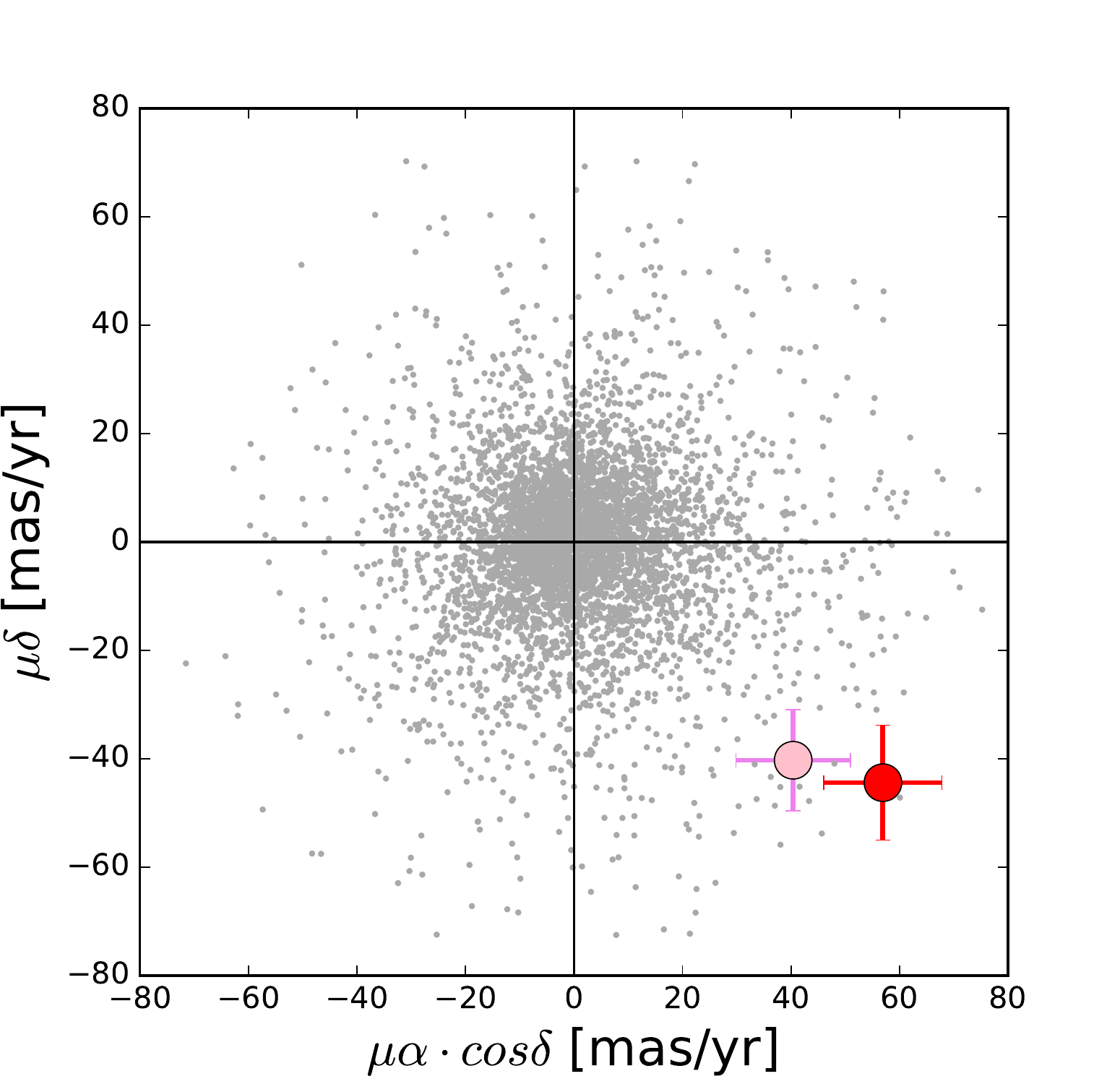}
   
   \caption{Proper motion diagram for 2MASS J0249$-$0557 AB (light pink) and 2MASS J0249$-$0557 c (red) from the VHS--2MASS cross correlation. Other cross-correlated objects in the same field of view are marked in grey.}
              \label{fig:PM}%
    \end{figure}

  \begin{figure}
   \centering
   \resizebox{\hsize}{!}{\includegraphics{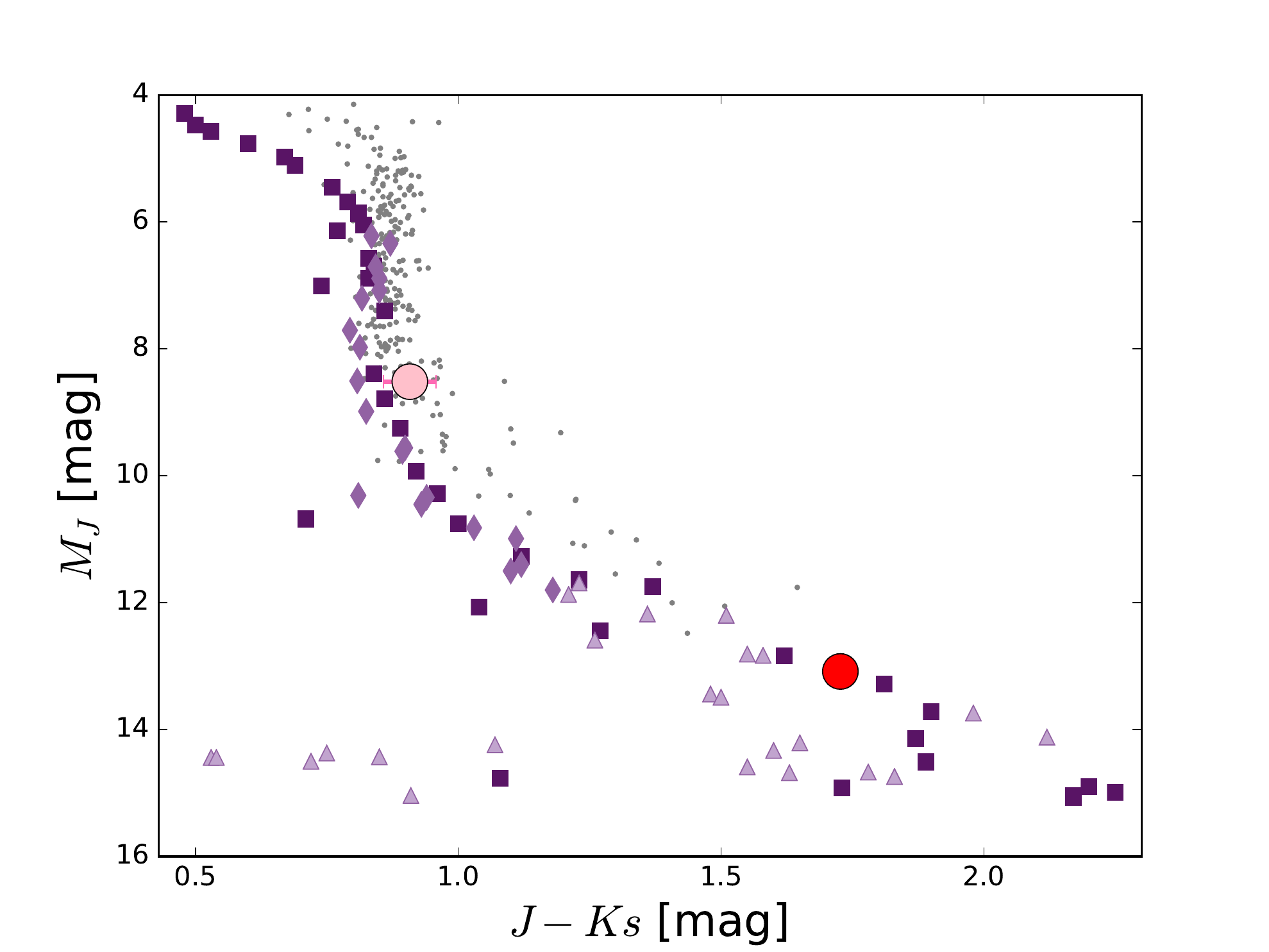}}
   \caption{$M_{J}$ vs. \textit{J$-$Ks} colour-magnitude diagram for 2MASS J0249$-$0557 AB (light pink) and 2MASS J0249$-$0557 c (red).   
 Other known $\beta$ Pic members are plotted as grey dots. 
 The photometric sequence of field dwarfs derived by  \citet[squares]{PecautMamajek2013}, \citet[triangles]{Dupuy2012}, and \citet[diamonds]{Lodieu2014} is also shown in purple. }
              \label{fig:JmKs_colormag}%
    \end{figure}
 
 \section{Independent identification of 2MASS~J0249$-$0557\,c}

As part of our efforts to characterise the frequency of wide substellar companions around young nearby stars, we performed a search for common proper motion companions to known members and candidate members of the $\beta$ Pictoris YMG, among other young stellar kinematic groups, using data from the Visible and Infrared Survey Telescope for Astronomy (VISTA) Hemisphere Survey \citep[VHS, ][]{McMahon2013} catalogue in combination with the Two Micron All-Sky Survey \citep[2MASS, ][]{2MASSpaper} catalogue. 

For this purpose, we first compiled a list of 348 known members and candidate members of $\beta$ Pictoris, which we obtained from \citet{Schlieder2010, Schlieder2012}, \citet{Malo2013}, \citet{Bell2015}, \citet{Best2015}, \citet{Gagne2015VII}, \citet{Gagne2015V}, \citet{Elliott2016}, \citet{Messina2017}, \citet{Schneider2017}, \citet{Shkolnik2017} and  \citet{Gagne2018XIII}. One of the members in this list is 2MASS~J0249$-$0557.

Secondly, we queried the VHS database to download all the available data in a circular area corresponding to 50\,000 AU around each $\beta$ Pic candidate member in the compilation. The angular area corresponding to this radius was calculated in most of the cases using the available parallaxes from $Gaia$ DR2. For the objects without parallaxes, we downloaded all the data in a 83.4\arcmin \,radius, which corresponds to a physical separation of 50\,000 AU for an object placed at 10 pc distance, considering this as a lower limit for the heliocentric distance of $\beta$ Pic members.

Then, we cross-matched the VHS data with the 2MASS catalogue to identify nearby objects whose proper motions matched those of the primaries within an error of 40 mas yr$^{-1}$ in total. This value is at least twice the typical dispersion in the measured proper motions of non-moving background objects, which is 10--20 mas yr$^{-1}$, depending on the time baseline between the VHS and 2MASS observations (9--19 years). To the targets with compatible proper motions, we later applied photometric selection criteria using VHS and 2MASS \citep{2MASSpaper} photometry and also DENIS and/or AllWISE data \citep{DENISpaper, WISEpaper} when available. We selected as good candidate companions those objects whose magnitudes and colours followed a well-defined photometric sequence with their primaries in different colour-magnitude diagrams, and which therefore can be placed at the same heliocentric distance. We also used $Gaia$ DR2 \citep{Gaiapaper, Gaiapaper2} astrometry for the brightest candidates to discard chance alignments.

As a result of this search, we found 2MASS~J0249$-$0557\,c,  placed at a distance of 39.947$\pm$0.014 arcsec from 2MASS~J0249$-$0557, with a position angle of 228.$^{\circ}$644 $\pm$ 0.$^{\circ}$020. These measured separation and position angles are compatible with those published in \citet{Dupuy2018}. This is the faintest candidate companion identified in our search. However, because we are limited to the completeness of 2MASS ($J$$\sim$16) in this search, other candidate companions with similar masses and spectral types may have remained undetected. Figure \ref{fig:finding_chart} shows the finding chart for the 2MASS~J0249$-$0557 system. The proper motion of the candidate companion agreed well with the motion of the primary, and its magnitude and colours were consistent with an L-type object located at the same heliocentric distance as the primary brown dwarf pair.  Figures \ref{fig:PM} and \ref{fig:JmKs_colormag} show the proper motion diagram and \textit{J}--\textit{J-Ks} colour magnitude diagram for the system.

2MASS~J0249$-$0557 was imaged by VHS, 2MASS, Pan-STARRS \citep{PanSTARRSpaper}, the Sloan Digital Sky Survey \citep[SDSS;][]{SDSSpaper}, WISE, and Gaia. The available photometric data for the system are presented in Table \ref{Table:gendata}.

   \begin{table}
   \tiny
      \caption[]{General data of 2MASS J0249$-$0557 AB and c.}
         \label{Table:gendata}
     $$ 
         \begin{array}{p{0.3\linewidth}lcc}
            \hline
            \hline
            \noalign{\smallskip}
            Astrometry &&  \mathrm{AB} &  \mathrm{c} \\
            \noalign{\smallskip}  
           \hline
            \noalign{\smallskip}
             R.A. (J2000) && 02:49:56.39  &  02:49:54.36    \\
           DEC (J2000) &&   -05:57:35.2   &  -05:58:01.5   \\
            Parallax (mas) \tablefootmark{a}  && 20.5\pm2.1 &     20.1\pm3.5      \\ 
           Distance (pc) \tablefootmark{a}    &&  \multicolumn{2}{c}{48.9^{+4.4}_{-5.4}}    \\ 
             Separation (arcsec) \tablefootmark{a} && 0.0444\pm0.0002 \arcsec & 39.959\pm0.005 \arcsec  \\
            Separation (AU) \tablefootmark{a} & & 2.17\pm0.22 &1950\pm200  \\ 
                  P.M. (mas yr$^{-1}$) \tablefootmark{a} && (42.9, -30.2) &  (46.0, -32.0) \\
                     &&   \pm (2.0,1.8)   &  \pm (2.3,2.1)   \\
                    \noalign{\smallskip}
            \hline
             \noalign{\smallskip}
              Spectroscopy && & \\
        \noalign{\smallskip}
         \hline
            \noalign{\smallskip}
                  Spectral type  &&   \mathrm{M6}\, $\tablefootmark{b} $  & \mathrm{L2.5\pm0.5~OPT, L3\pm1~NIR}  \,$\tablefootmark{c}$ \\
                  Li pEW (\AA)   && 0.59\pm0.05\, $\tablefootmark{b}$ & \lesssim 5 \\
                  H$\alpha$ pEW (\AA)    && -11.61\pm0.1\, $\tablefootmark{b}$ & -90^{+20}_{-40}  \,\,$\tablefootmark{c}$\\
            \noalign{\smallskip}
            \hline
             \noalign{\smallskip}
              Photometry && & \\
        \noalign{\smallskip}
         \hline
            \noalign{\smallskip}
            $Gaia$ \textit{g}&&   15.6894 \pm 0.0012  &  -  \\
             \noalign{\smallskip}
             SDSS \textit{u}&&   21.47 \pm 0.14   &  -  \\
             SDSS \textit{g}&&   18.721 \pm 0.008   &  -  \\ 
                SDSS \textit{r}&&  17.197 \pm 0.006    &  -  \\ 
                SDSS \textit{i}&&   14.977 \pm 0.005   &  20.76 \pm 0.13  \\
                SDSS \textit{z}&&  13.754 \pm 0.005    &  19.28 \pm 0.15  \\
                                  \noalign{\smallskip}
                Pan-STARRS \textit{g}&& 18.419 \pm 0.008     &  - \\
                Pan-STARRS \textit{r}&& 17.145 \pm 0.005     &  - \\
                Pan-STARRS \textit{i}&& 14.992 \pm 0.004     &  21.36 \pm 0.06 \\
                Pan-STARRS \textit{z}&&  13.984 \pm 0.004    &  19.92 \pm 0.03\\
                Pan-STARRS \textit{y}&&  13.409 \pm 0.004    &  18.93 \pm 0.03  \\
                \noalign{\smallskip}
         VHS \textit{J} &&  11.9667 \pm 0.0007   &  16.531 \pm 0.008  \\
                 VHS \textit{Ks} &&  11.2535 \pm 0.0009   &  14.804 \pm 0.009  \\
                \noalign{\smallskip}             
                  2MASS \textit{J}&&  11.96 \pm  0.03   &  16.55 \pm 0.11  \\
                  2MASS \textit{H}&&   11.36 \pm 0.03   &  15.48 \pm 0.13  \\
                  2MASS \textit{K}&&   11.06 \pm 0.02   &  14.88 \pm 0.12  \\
                
                \noalign{\smallskip}     
                 WISE \textit{w1}&&   10.84 \pm 0.02   &  14.13 \pm 0.03  \\
                 WISE \textit{w2}&&   10.60 \pm 0.02   &  13.59 \pm 0.04  \\
                 WISE \textit{w3}&&   10.39 \pm 0.06   &  -  \\

        \noalign{\smallskip}
            \hline
            \hline

         \end{array}
     $$ 
     \tablefoot{
\tablefoottext{a}{Data from \citet{Dupuy2018}.}
\tablefoottext{b}{Data from \citet{Shkolnik2017}.}
\tablefoottext{c}{Data from this work.}
}
   \end{table}

%
\begin{table*}
\setlength\tabcolsep{4.5pt}
\caption{Observation log for 2MASS J0249$-$0557 c}             
\label{table:obslog}      
\centering          
\begin{tabular}{c c c c c c c c c c c c}     
\hline\hline       
Obs. Date & Telesc/Instrum & Mode & Grating/ & Wavelength & Slit & Pl. Scale & Disp. & Res. & Exp. & Airmass & Seeing \\ 
 &  & & Filter & Range [$\mu$m] &[\arcsec] & [\arcsec/pix] & [$\AA /pix$] & Power & Time [s] &  & [\arcsec] \\ \hline 
 \noalign{\smallskip}                   
11 Sep 2016  & VISTA/VIRCAM  & Img & J, Ks & -- & -- & 0.34 & -- & -- & 8$\times$15, 8$\times$7.5 & 1.1 & 0.9 \\
21 Jun 2018  & NTT/SofI & Spec & GR & 1.53--2.52 & 1 & 0.29 & 10.22 & 600 & 4$\times$300 & 1.4--1.5  & 0.9 \\ 
21 Jun 2018  & NTT/SofI & Spec & GB & 0.95--1.64 & 1  & 0.29 & 6.96 & 600 & 4$\times$600 & 1.6--1.8  & 0.9 \\
4 Nov 2018  & NTT/SofI & Spec & GR & 1.53--2.52 & 1 & 0.29 & 10.22 & 600 & 4$\times$300 & 1.2--1.3  & 0.6 \\ 
4 Nov 2018  & NTT/SofI & Spec & GB & 0.95--1.64 & 1  & 0.29 & 6.96 & 600 & 4$\times$600 & 1.1--1.2  & 0.6 \\  
  
27 Jan 2019  & GTC/OSIRIS & Spec & R500R & 0.48--1.00 & 0.8 & 0.25 & 4.88 & 440  & 4$\times$1600 & 1.2--1.4 & 0.7 \\
   &  &  &  &  &  &  & &  &  &  &  \\    
\hline  
\hline                
\end{tabular}
\end{table*}

\section{Follow-up observations and data reduction}

\subsection{GTC/OSIRIS optical spectroscopy}

2MASS J0249$-$0557 c was observed on 27 January 2019 with the Optical System for Imaging and low-intermediate-Resolution Integrated Spectroscopy \citep[OSIRIS; ][]{OSIRISpaper} spectrograph at the 10.4 m Gran Telescopio Canarias (GTC) in La Palma, Canary Islands. The observations were performed in long-slit spectroscopy mode, using the R500R grism (wavelength coverage 0.48--1.00 $\mu$m) and the 0.8\arcsec wide slit placed in parallactic angle, providing a resolution of $\sim$440 at 7200~$\AA$. Four exposures of 1600s were obtained using an ABBA nodding pattern with a 14\arcsec offset along the slit. The object acquisition was performed using the Sloan \textit{z} filter. The seeing value was 0.7\arcsec. The spectrophotometric standard G191-B2B (DA white dwarf, $V$=11.7) was also observed with the same instrumental configuration to correct for the instrumental response. The observation log is shown in Table \ref{table:obslog}.

The data reduction was performed using standard routines within the Image Reduction and Analysis Facility (IRAF) environment \citep{IRAFpaper1, IRAFpaper2}. The images were corrected for bias and were flat-fielded. Then, the 2D images of the four individual exposures of the target were corrected for bad pixels and cosmic rays using IMEDIT. The four target spectra and the spectrophotometric standard spectrum were extracted using the APALL routine. The spectra were wavelength calibrated at each nodding position using HgAr, Ne, and Xe arc images. The spectrum of the standard was corrected for second-order contamination of the light at wavelengths 4800--4900~\AA \,using a spectrum taken with the $z$-band filter, following the procedure explained in \citet{Zapatero2014}. We obtained the sensitivity function from the standard star spectrum. Then, the spectra of the target were corrected for instrumental response using this sensitivity function. Finally, for an optimal combination of the four spectra, the data were aligned in wavelength with respect to the first exposure by cross-correlating the observations (the regions with strong telluric absorption were excluded from the cross correlation). The final combined spectrum is shown in Figures \ref{fig:OPT_vsKirk} and \ref{fig:OPT_vsYoung}.

\subsection{NTT/SofI infrared spectroscopy}

We obtained near-infrared low-resolution spectroscopy of 2MASS J0249$-$0557 c on 21 June 2018 and 4 November 2018 using the Son of Isaac (SofI) spectrograph \citep{1998Msngr..91....9M} mounted on the 3.6 m New Technology Telescope (NTT) at the La Silla Observatory, Chile, to determine the spectral type of the candidate companion and search for features of youth. We used long-slit spectroscopy mode with the 1.0\arcsec \,wide slit, and the low-resolution blue and red grisms (wavelength ranges 0.95--1.64 $\mu$m and 1.53--2.52 $\mu$m, respectively, resolving power 600, and nominal dispersion 6.96 and 10.22 $\AA$ pix$^{-1}$, respectively). The exposures were taken using an ABBA nodding pattern along the slit, with 600s and 300s individual exposure times in each position for blue and red grisms, respectively. We also observed the B7 telluric standards HIP14143 and HIP17457 after the exposures and at similar airmasses to correct for telluric absorption. To perform the flat-field correction and wavelength calibration, we also acquired continuum-lamp images and Xe arc-lamp images. A log of these observations is included in Table~\ref{table:obslog}.

The data reduction was performed using the SofI pipeline by ESO within the Gasgano environment \citep{Gasgano} and standard IRAF routines for both the target and telluric standards. The raw spectroscopic images were flat-field and dark corrected, aligned, and combined using Gasgano. We then used the IRAF APALL routine to extract the spectra from these 2D combined images, and wavelength calibrated them using the Xe arcs. We manually removed the intrinsic lines of the spectrum of the telluric star. Finally, to correct for the instrumental response, the target spectra were divided by the telluric spectra and multiplied by a black body of 14\,500 K, which is the corresponding temperature for the telluric star spectral type \citep{PecautMamajek2013}. For a better result in the signal-to-noise ratio, we combined the resulting spectra from the two observations. The final spectrum is shown in Figures \ref{fig:NIR_subplots} and \ref{fig:NIR_subplots_young}.

\section {Results and discussion}

\subsection{Optical characterisation and spectral classification}

\begin{figure}
   \centering
   \includegraphics[width=9cm]{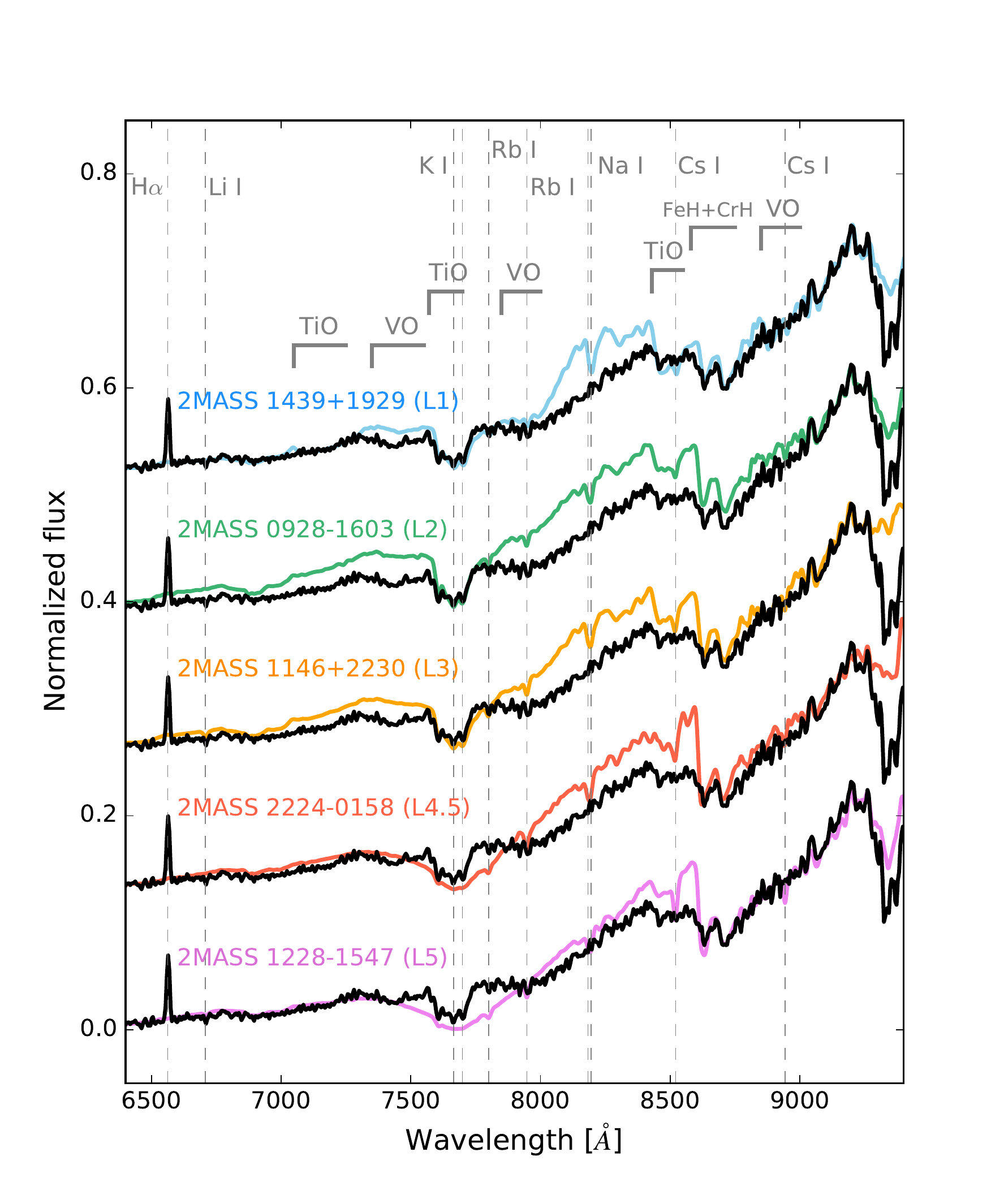}
   \caption{GTC/OSIRIS optical spectrum of 2MASS J0249$-$0557 c (black) compared to old field optical L standards from \citet{Kirkpatrick1999} and \citet{Kirkpatrick2000} observed with Keck-I/LRIS (coloured). LRIS spectra were convolved with a Gaussian to match the OSIRIS spectrum resolution. Spectra were normalised at wavelength $\sim$9200 $\AA$ and shifted by a constant for clarity. Several spectral features are marked in grey.}
              \label{fig:OPT_vsKirk}%
    \end{figure}

We have compared the optical spectrum of 2MASS J0249$-$0557 c with several young and old L-dwarf templates to determine the optical spectral type and analyse its spectral features. Figure \ref{fig:OPT_vsKirk} shows the comparison between the GTC/OSIRIS optical spectrum of 2MASS J0249$-$0557 c and some old field L-type standards from \citet{Kirkpatrick1999} and \citet{Kirkpatrick2000}, observed with the Low Resolution Imaging Spectrometer (LRIS) and the 400/8500 grism (R$\sim$1100) on the Keck-I Telescope. LRIS spectra were convolved with a Gaussian to match the GTC/OSIRIS spectral resolution. 

From this comparison, we find that the general spectral energy distribution of 2MASS~J0249$-$0557\,c in the optical fits that of the early and intermediate old-field L dwarfs. When we focus on the pseudo-continuum, the overall shape of the spectrum, particularly above 8000~$\AA$, coincides with that of the intermediate old field dwarf of L5 spectral type. However, the depth of the VO, TiO, FeH, and CrH absorption bands in the optical spectrum of the target are better reproduced by early old field L dwarfs. These differences in the spectral features can be explained by the effect of the low gravity of the young object, which is still contracting. These discrepancies are frequent when young objects are compared with old counterparts, and can be misinterpreted as a difference in spectral type. For example, the shapes of the VO bands at 7300--7600~$\AA$ and 7800--8000~$\AA$ are closer in depth to the L1--L2 spectral templates. TiO at 8400--8600~$\AA$ is weakened and closer to later spectral types. This can also be seen at the 7000--7400~$\AA$ TiO band. Alkali lines are also weakened, as can be seen in the NaI doublet at 8183--8195~$\AA$ or in the CsI line at 8521~$\AA$. 

We also observed a tentative detection of the absorption line of Li I at 6708~$\AA$, with a pEW of $\lesssim$ 5~$\AA$. Nevertheless, the feature is barely distinguishable from the spectral noise at a 2$\sigma$ level. The obtained value is compatible with a total preservation of lithium, as expected for a $\beta$ Pictoris L2--3 member \citep{Kirkpatrick2000}.

\begin{figure}
   \centering
   \includegraphics[width=9cm]{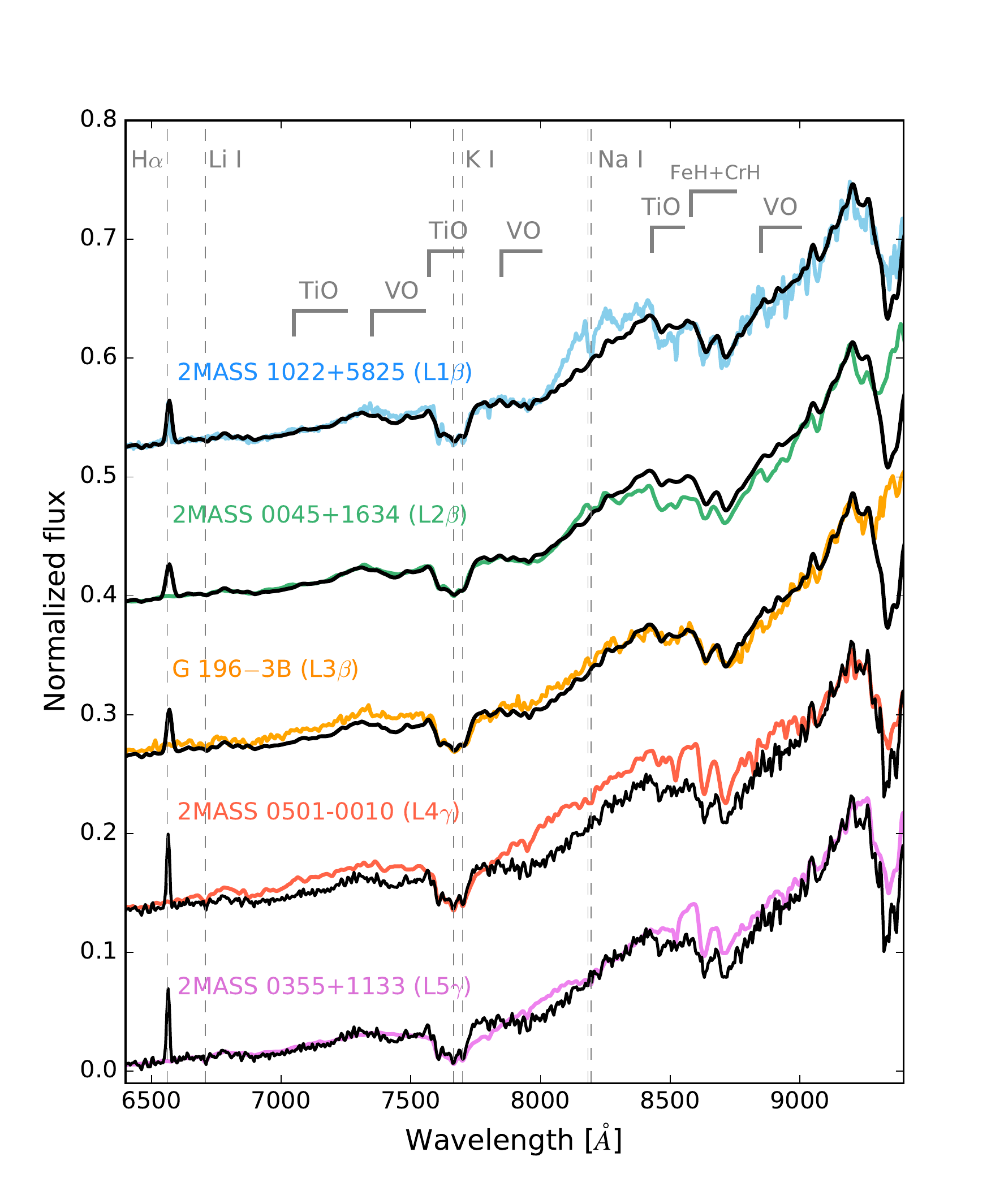}
   \caption{GTC/OSIRIS optical spectrum of 2MASS J0249$-$0557~c (black) compared to other known young L objects (coloured) from \citet[][2MASS 1022+5825, 2MASS 0501$-$0010 and 2MASS 0355+1133]{Cruz2009} and \citet[][2MASS 0045+1634 and G196--3B]{Zapatero2014}. The GTC/OSIRIS spectrum was convolved with Gaussians to match the young L1--L3 template spectra resolutions. For the L4--L5 comparisons, the template spectra were convolved to match the target spectral resolution. Spectra were normalised at wavelength $\sim$9200 $\AA$ and shifted by a constant for clarity. Some spectral features are marked in grey. }
              \label{fig:OPT_vsYoung}%
    \end{figure}

For a more accurate spectral type classification, we also compared the optical spectrum of 2MASS J0249$-$0557~c to the spectra of some known young objects with spectral types L1--L5. Figure \ref{fig:OPT_vsYoung} shows these comparisons. This figure shows the strong resemblance of the spectrum of 2MASS J0249$-$0557~c to the optical spectra of 2MASS J00452143+1634446 (2MASS 0045+1634), which is classified as an L2, and G~196--3B \citep{Rebolo1998}, which is classified as an L3. 2MASS 0045+1634 is a young substellar object, with an estimated age of 10--100 Myr \citep{Zapatero2014}. It has been claimed as a likely member of the Argus moving group \citep[40--50 Myr;][]{Torres2008, Zuckerman2019} by \citet{Gagne2014}. G~196--3B is a young object, with an estimated age of 20--300 Myr, and it is also likely younger than 100 Myr \citep{Zapatero2010, Zapatero2014}. \citet{Gagne2014} indicated a moderate probability that G196--3B belongs to the AB Doradus moving group \citep[$\sim$125 Myr;][]{Zuckerman2004, Luhman2005, Barenfeld2013}. The spectra coincide in the overall shapes and in the depth of the oxide bands, suggesting that they may have  similar temperatures and gravities. The most remarkable difference between our target and these comparison spectra in the optical is the H$\alpha$ emission line, which is strong in 2MASS~J0249$-$0557~c but it is not detected in either 2MASS 0045+1634 or G~196--3B \citep{Mohanty2003}. We discuss this in more detail in section \ref{subsection:halpha}. 

We have also computed spectral indices from \citet{Martin1999} to assess the spectral classification of this target. They are listed in Table \ref{table:indices}. We excluded other commonly used indices related to the oxides because these features are strongly affected by gravity and are not adequate for spectral type determination in young objects. Based on our comparison with young and field L dwarfs and the spectral indices, we finally adopted a classification in the optical of L2.5$\pm$0.5.

\begin{table}
\setlength\tabcolsep{10pt}
\caption{Spectral indices for spectral type determination.}             
\label{table:indices}   
 \begin{tabular}{lcll}
 \hline \hline
\noalign{\smallskip}                    
Index & Value & SpT & Index Ref.\tablefootmark{a}\\ 
\noalign{\smallskip}
\hline\hline       
  \noalign{\smallskip}

Optical &  &  &  \\ 
\noalign{\smallskip}
\hline
 \noalign{\smallskip} 
 PC3 & 2.78 & L1.5$\pm$0.5 & M99 \\
 PC6 &  21.9  & M8--L4   & M99 \\
\noalign{\smallskip}
 \hline
\noalign{\smallskip}                     
Infrared &  &  &  \\ 
\noalign{\smallskip}
\hline
 \noalign{\smallskip} 
 H$_{2}$O \tablefootmark{b} & 1.35  & L4$\pm$0.5 & A07 \\
  \noalign{\smallskip} 
 H$_{2}$OD \tablefootmark{b} &  0.99 & L0.25$\pm$0.75 & ML03 \\
  \noalign{\smallskip} 
 H$_{2}$O-1 \tablefootmark{b} & 0.59  & L3.25$\pm$1.00  & S04 \\
  \noalign{\smallskip} 
 H$_{2}$O-2 \tablefootmark{b} & 0.82  & L1.25$\pm$0.50  & S04 \\  
 \noalign{\smallskip}

\hline  

 \hline

\end{tabular}

\tablefoot{
\tablefoottext{a}{References: A07 -- \citet{Allers2007}, M99 -- \citet{Martin1999}, ML03 -- \citet{McLean2003}, S04 -- \citet{Slesnick2004}.}
\tablefoottext{b}{Results obtained using the \citet{Allers2013} polynomial fits.}
} 
\end{table}

\nocite{Wilking2005; Slesnick2006; Kirkpatrick1999; Reid1995; Martin1999}

\subsection{Infrared spectral classification}

\begin{figure}
   \centering
   \includegraphics[width=9.7cm]{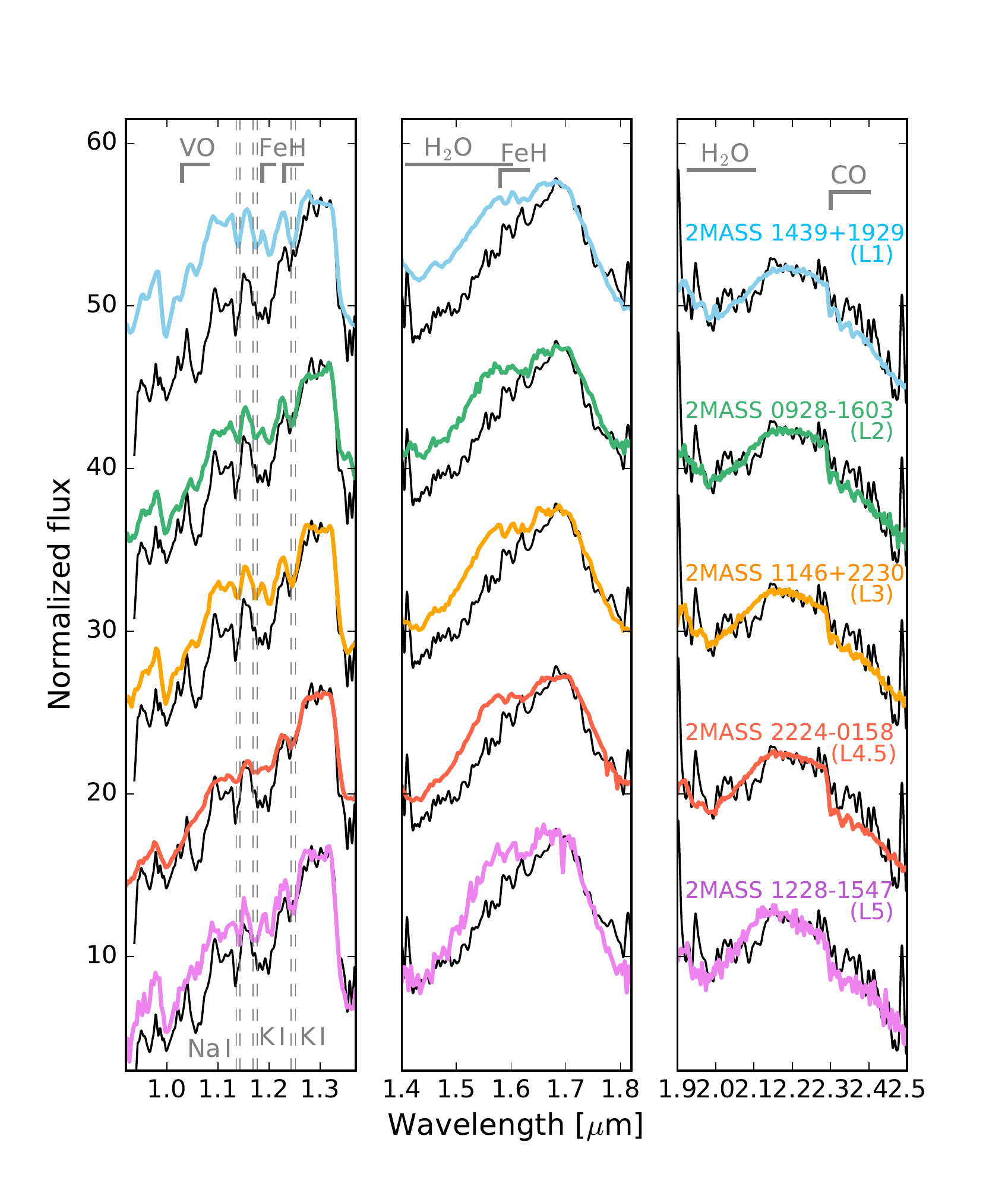}
   \caption{NTT/SofI spectrum of 2MASS J0249$-$0557~c (black) compared to old field L standards from \citet{Kirkpatrick1999} and \citet{Kirkpatrick2000} observed with IRTF/SpeX (coloured) by \citet{Burgasser2004B}, \citet{Burgasser2010A}, and \citet{Cruz2018}. The NTT/SofI spectrum was convolved with a Gaussian to match the resolution of SpeX spectra. \textit{J}, \textit{H,} and \textit{K} bands are displayed and normalised separately. Some spectral features, such as the VO, FeH, H$_{2}$O, and CO bands or the NaI and KI doublets, are marked in grey.}
              \label{fig:NIR_subplots}%
    \end{figure}

To determine the near-infrared spectral type, we compared our NTT/SofI spectrum with infrared spectra of known young and field objects. Figure \ref{fig:NIR_subplots} shows a comparison of our 2MASS J0249$-$0557~c near-infrared spectrum with several old field L dwarfs from \citet{Kirkpatrick1999} and \citet{Kirkpatrick2000}. This figure shows that the spectral energy distribution of our target in the near-infrared bands resembles that of the L3 and L4.5 spectral types, except for the different shapes of the $H$ band and the depth of some molecular bands such as VO and FeH, which are more intense in 2MASS~J0249$-$0557~c. The triangular shape of the $H$ band \citep[e.g.,][]{Lucas2001, Allers2013} is a remarkable feature that reveals the young nature of this object.

\begin{figure}
   \centering
   \includegraphics[width=9.6cm]{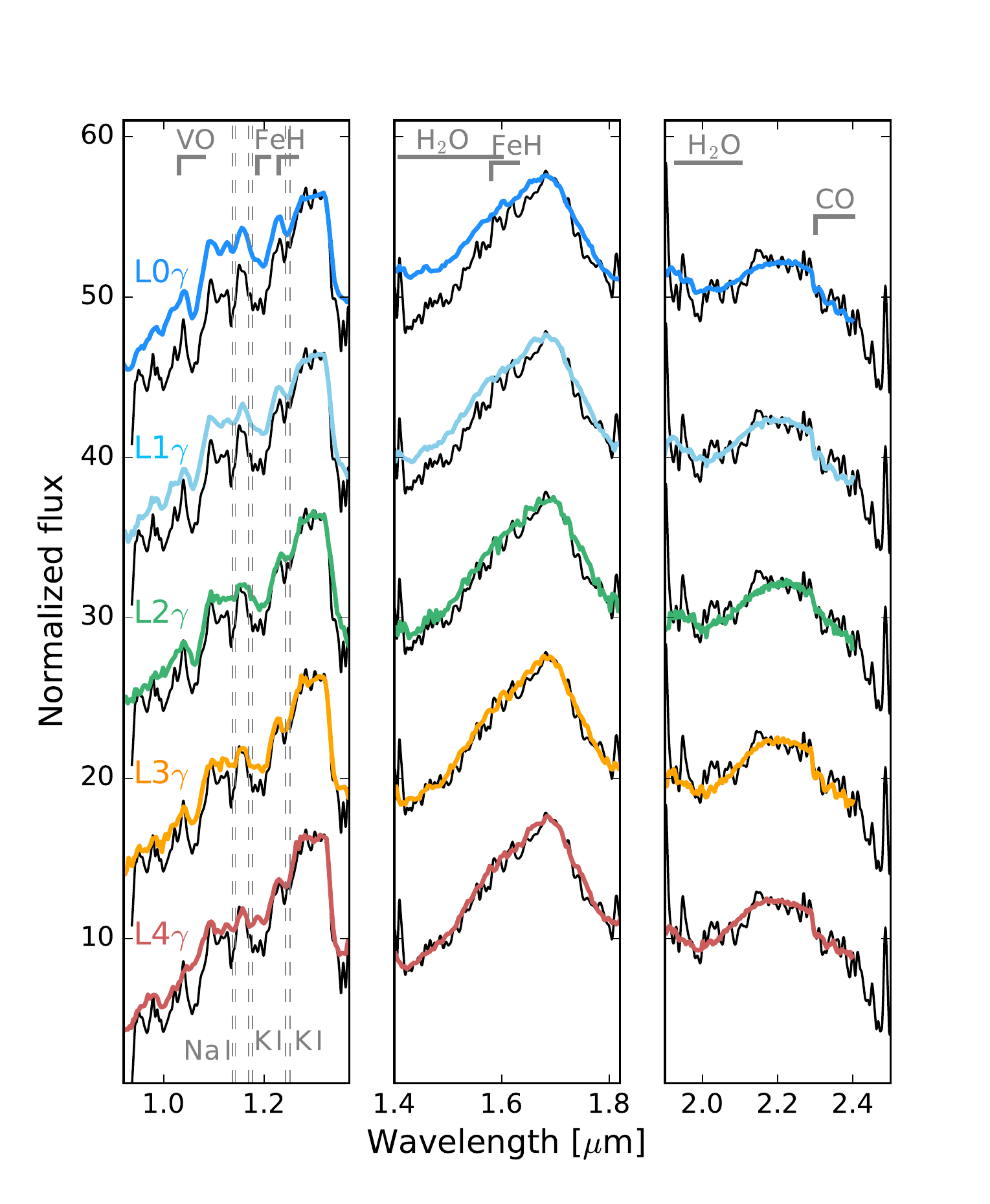}
   \caption{Same as Fig. \ref{fig:NIR_subplots}, but comparing the NTT/SofI spectrum of 2MASS J0249$-$0557 c (black) to near-infrared very low gravity L dwarf templates from \citet{Cruz2018} (coloured). The NTT/SofI spectrum was convolved with a Gaussian to match the spectral resolution of the templates.}
              \label{fig:NIR_subplots_young}%
    \end{figure}

Figure \ref{fig:NIR_subplots_young} shows the near-infrared spectrum of 2MASS J0249$-$0557~c compared to very low gravity L spectral templates from \citet{Cruz2018}. The spectrum fits the L2$\gamma$, L3$\gamma,$ and L4$\gamma$ templates in each of the individual near-infrared bands rather well. This is clearly visible in the slopes of the H$_{2}$O absorption bands at either side of the $H$ band, and also at the red side of the $J$ band. However, there are some discrepancies. For example, the VO band at 1.06~$\mu$m is stronger in 2MASS J0249$-$0557~c than in the templates. For an old object this would imply an earlier spectral type, but in this case, it may indicate a lower gravity, and hence a younger age for 2MASS J0249$-$0557~c compared to the templates. Although all the objects used to produce these template spectra show very low gravity features, and at least one of the three objects used to produce the L3$\gamma$ template is probably a member of $\beta$ Pic \citep[see][]{Cruz2018}, they may have a wider range of ages and many of them may be older than this young moving group.
  
  To confirm this spectral type classification, we also estimated the near-infrared spectral indices based on H$_{2}$O absorption bands from \citet{Allers2007}, \citet{McLean2003} and \citet{Slesnick2004} and the best  gravity-dependent indices defined by \cite{Lodieu2018} for the gravity type. These indices are shown in Table \ref{table:indices} and Table \ref{table:lgindices}. Based on these comparisons and the spectral indices, we finally adopted a spectral type of L3$\pm$1 $\gamma$ in the near-infrared for 2MASS~J0249$-$0557~c, where the $\gamma$ suffix is an indication of very low gravity features. This classification is compatible with the spectral type determination of L2$\pm$1 VL-G assigned by \citet{Dupuy2018}, where VL-G also indicates a very low gravity. This is consistent with the signs of youth we have found in the optical and near-infrared spectra and with the expected age for a member of $\beta$ Pic.

\begin{table}
\begin{center}
\caption{Gravity-dependent indices.}             
\label{table:lgindices}   
 \begin{tabular}{p{0.14\linewidth}lccl}
 \hline
\noalign{\smallskip}                    
Index & Value & Gravity\tablefootmark{a} & Index Ref.\tablefootmark{b} \\ 
\hline\hline

 \noalign{\smallskip} 
 H-cont & 0.99  & $\gamma$  & AL13 \\
 \noalign{\smallskip}
 CH$_{4}$-H & 1.16  & $\beta$--$\gamma$  & B06 \\
 \noalign{\smallskip}
 VO$_{z}$ & 1.50  & $\gamma$ & AL13 \\
 \noalign{\smallskip}
 H$_{2}$O-K & 0.86  & $\beta$--$\gamma$ & B06 \\

\hline  
 
 \hline              
   
\hline                  
\end{tabular}
\end{center}
\tablefoot{
\tablefoottext{a}{Gravity indices as described in \citet{Kirkpatrick2005} and \citet{Cruz2009}.} Assigned following the plots in Fig. A1 of \citet{Lodieu2018}.
\tablefoottext{b}{References: AL13 -- \citet{Allers2013}, B06 -- \citet{Burgasser2006}}
} 

\end{table}

\subsection{Comparison between optical and near-infrared spectral classification}

We adopted a spectral type of L3$\pm$1 in the near-infrared, which agrees well with previous determination by \citet{Dupuy2018}, and L2.5$\pm$0.5 in the optical. This is the first classification of this object in this wavelength range. Although the uncertainty in the spectral type determination of young objects is larger than in older field L dwarfs because the low-gravity templates do not constitute a homogeneous sample of targets with the same age and hence evolutionary state, our classification in the optical and in the near-infrared is compatible. 

Previous studies in very young open clusters and associations with ages younger than $\sim$\,10\,Myr have shown that L dwarfs may have a slightly earlier spectral type in the optical than in the near-infrared \citep{Zapatero2017, Lodieu2018}. 
However, this does not seem to be the case for L dwarfs belonging to YMGs and associations at ages $>$10\,Myr \citep{Cruz2018}. The $\beta$ Pic group is close to the frontier between the two age groups and therefore is a favourable place to investigate this issue in more detail.

\subsection{H$\alpha$ emission in 2MASS~J0249$-$0557~c \label{subsection:halpha}}

As pointed out in previous sections, the strong H$\alpha$ emission line at 6563~$\AA$ is a remarkable feature of this object. To investigate the chromospheric activity and its variability, we measured the H$\alpha$ pEW in the final combined GTC/OSIRIS spectrum by fitting a Gaussian profile to the line using the SPLOT task within the IRAF environment. We repeated this procedure several times, selecting a wide range of acceptable continuum levels. The average measured pEW is $-90^{+20}_{-40}\AA$. The quoted error takes the spread of the measured pEWs with the different choices of the continuum level into account.

This object was also observed by the Extended Baryon Oscillation Spectroscopic Survey \citep[eBOSS; ][]{eBOSSpaper} from the SDSS on 3 December 2015. This survey is carried out using the BOSS spectrographs \citep{BOSSspectrograph} mounted on the SDSS 2.5m telescope \citep{Gunn2006} at the Apache Point Observatory, New Mexico. The eBOSS optical spectral coverage is 3600--10400 $\AA$, and the spectral resolution is 1300--2500 \citep{Alam2015}. The eBOSS spectra are publicly available in the SDSS Sky Server.\footnote{\url{http://skyserver.sdss.org/dr15/en/home.aspx}} Due to the faintness of the object in the bluer range, the signal-to-noise ratio of the spectrum is rather poor in the H$\alpha$ region, and the pseudo-continuum level is nearly zero. Nevertheless, the H$\alpha$ emission line can be recognised. We measured a lower limit in the H$\alpha$ emission of 20 $\AA$ from the eBOSS spectrum (pEW $< -20\AA$).

\begin{figure}
   \centering
   \includegraphics[width=4.12cm]{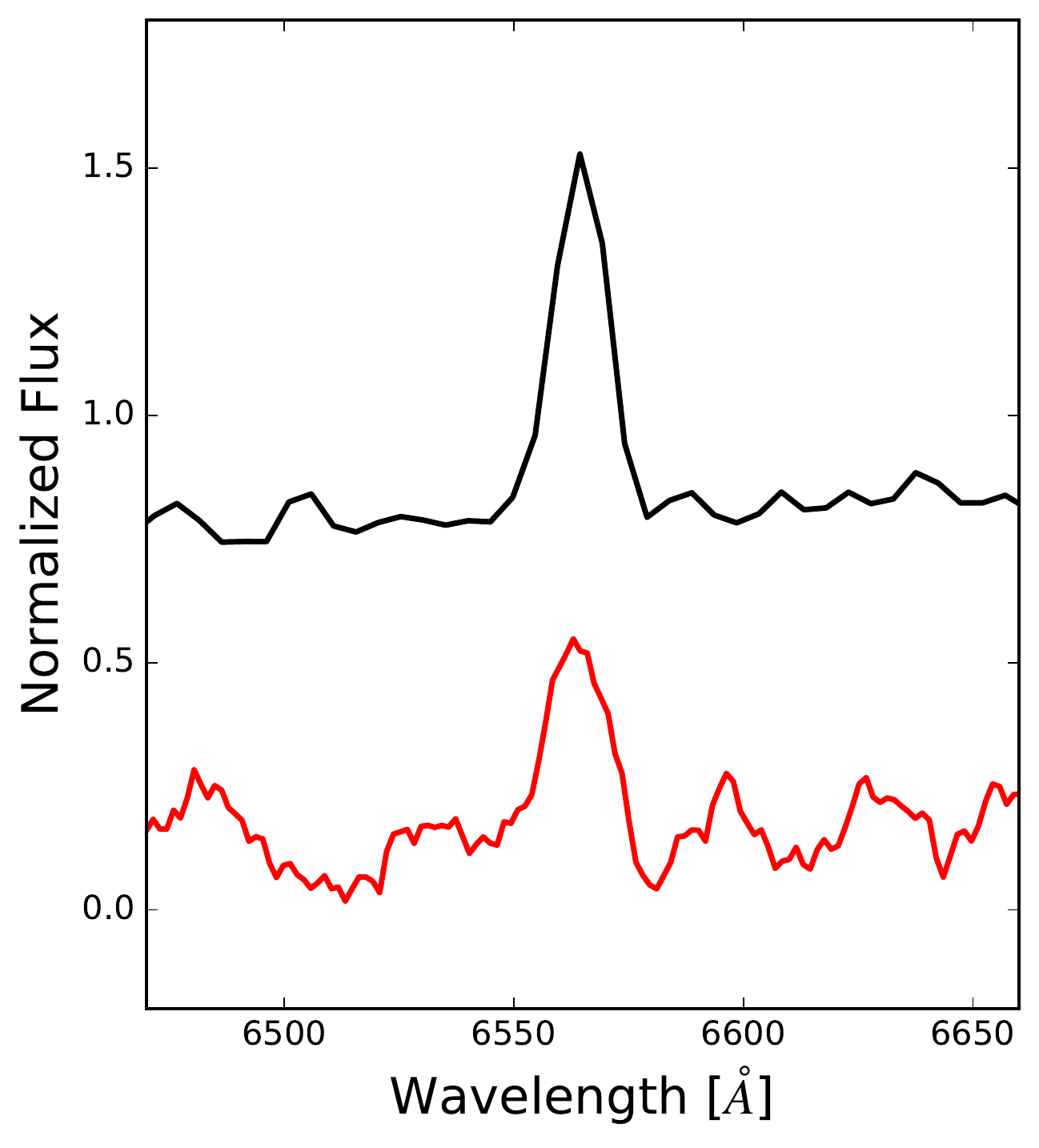}
   \includegraphics[width=4.7cm]{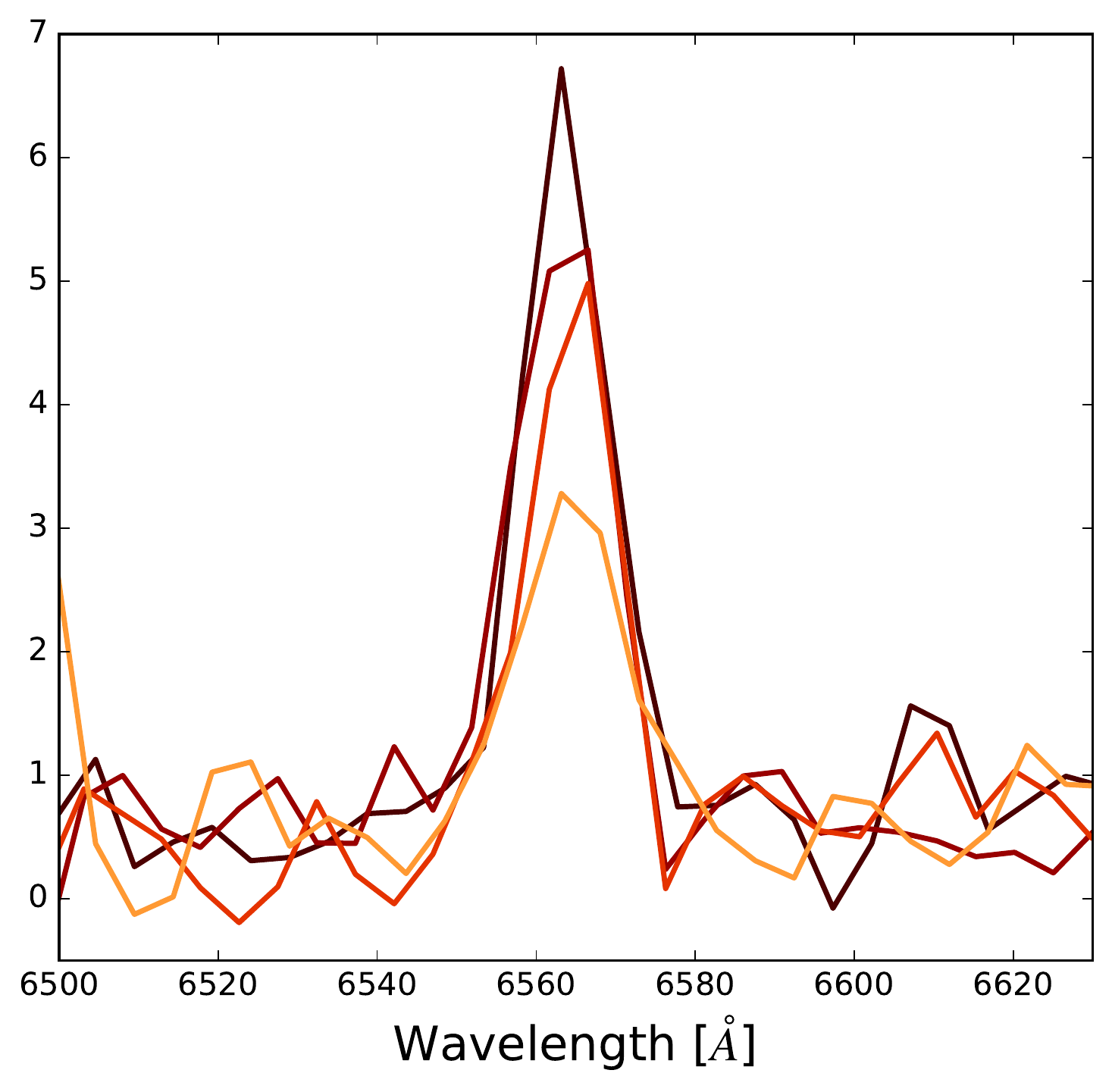}
   \caption{Variation in the H$\alpha$ emission line. The left panel shows the H$\alpha$ line in our GTC/OSIRIS spectrum (black) compared with the eBOSS spectrum (red). The eBOSS spectrum has been smoothed for a better visualisation. The right panel shows the H$\alpha$ line extracted from the four GTC/OSIRIS individual exposures separately.}
              \label{fig:Halpha_sloan+indiv}%
    \end{figure}

The left panel of Figure \ref{fig:Halpha_sloan+indiv} shows the comparison between the H$\alpha$ lines of the eBOSS and the GTC/OSIRIS spectra of 2MASS J0249$-$0557~c. The time baseline between the observations is three years. This figure shows that the emission of H$\alpha$ appears at both epochs. Although it is possible that we have observed the object during two isolated flare episodes by chance, it is therefore more likely that the H$\alpha$ emission in 2MASS~J0249$-$0557~c is persistent on timescales of years and is not related to episodic events. 

We have also analysed the variations in the H$\alpha$ emission within the individual exposures of the GTC/OSIRIS observations. Four exposures of 1600s ($\sim$27 min) were taken. We extracted the individual spectra and measured the H$\alpha$ pEW in each one of them separately. Table \ref{table:halpha_indiv} shows the H$\alpha$ measurements of the individual exposures and of the combined spectrum. The right panel of Figure \ref{fig:Halpha_sloan+indiv} shows the comparison between the four individual exposures. We find consistent values of the H$\alpha$ pEW in them, with some variations that do not seem to be significant within the errors. Because the pseudo-continuum flux of the target in the wavelength region around H$\alpha$ is low, the measured errors of the pEW in the individual spectra are relatively large.

\begin{table}
\begin{center}
\setlength\tabcolsep{8pt}
\caption{H$\alpha$ measurements for 2MASS~J0249$-$0557~c.}             
\label{table:halpha_indiv}   
 \begin{tabular}{lll}
 \hline
\noalign{\smallskip}                    
 & MJD & H$\alpha$ pEW ($\AA$)    \\
\noalign{\smallskip} 
\hline
\noalign{\smallskip} 
\noalign{\smallskip} 
eBOSS & 57359 & $<-$20 \, \tablefootmark{a}    \\
\noalign{\smallskip} 
\noalign{\smallskip} 
OSIRIS Indiv. 1 & 58510.841 & $-$110$^{+40}_{-100}$    \\
\noalign{\smallskip} 
\noalign{\smallskip} 
OSIRIS Indiv. 2 & 58510.859 &  $-$110$^{+40}_{-80}$        \\
\noalign{\smallskip} 
\noalign{\smallskip} 
OSIRIS Indiv. 3 & 58510.878  & $-$130$^{+70}_{-60}$ \, \tablefootmark{b} \\
\noalign{\smallskip} 
\noalign{\smallskip} 
OSIRIS Indiv. 4 & 58510.897 & $-$80$^{+40}_{-70}$   \\
\noalign{\smallskip} 
\noalign{\smallskip} 
OSIRIS Comb. & 58510.87 & $-$90$^{+20}_{-40}$    \\

\noalign{\smallskip} 
\hline\hline       
                
\end{tabular}

\end{center}
\tablefoot{
\tablefoottext{a}{The pseudo-continuum emission in the H$\alpha$ region is too close to zero in this spectrum.}
\tablefoottext{b}{Very low continuum level in the H$\alpha$ region.}
} 

\end{table}

\subsection{H$\alpha$ emission in ultra-cool dwarfs}

The H$\alpha$ line found in emission in ultra-cool dwarfs is usually associated with chromospheric activity and/or accretion  \citep{Hawley1996, Gizis2000, White2003, Mohanty2003, Muzerolle2003, Jayawardhana2003, Mohanty2005, Caballero2006}. It is a common feature in late-M and early-L objects and occurs most frequently at $\sim$M7 spectral type. It decreases dramatically for mid-L and later spectral types \citep{Gizis2000, Schmidt2015, Pineda2016}. In addition, although some old L-type objects may present H$\alpha$ emission lines due to flares and close binarity, a strong and stable H$\alpha$ emission is commonly associated with youth \citep{Liebert2003}. However, many known young objects lack H$\alpha$ emission \citep[e.g.][]{Martin2010, Lodieu2018, Chinchilla2020}. Stable and strong H$\alpha$ emission therefore is a sufficient but not necessary indicator of youth.

We can find many examples of young late-M brown dwarfs with strong H$\alpha$ in emission in open clusters and associations of different ages \citep[e.g. ][]{Luhman1997, Bejar1999, Luhman1999, Ardila2000, Comeron2000, Martin2001, Briceno2002, Zapatero2002, Barrado2003, White2003, Mohanty2005, Lodieu2006, Slesnick2006, Barrado2007, Lodieu2011}. 
However, there are very few young L-dwarfs with a pronounced H$\alpha$ emission. An outstanding case is 2MASS J11151597+1937266, an L2$\gamma$ in the planetary mass regime with a strong H$\alpha$ emission \citep[pEW=560$\pm$82 $\AA$,][]{Theissen2017, Theissen2018}. In Sigma Orionis \citep[1--8 Myr;][]{Zapatero2002}, S Ori 71, a young L0 brown dwarf was reported to have an H$\alpha$ pEW of around $-$700 $\AA$ \citep{Barrado2002}. A similar case in this open cluster is S\,Ori\,55, a substellar M9 object at the deuterium-burning mass limit with a variable H$\alpha$ with pEW of 180--410 $\AA$ \citep{Zapatero2002_2}. In these two cases the presence of strong H$\alpha$ emission is attributed to the presence of an accretion disk, mass transfer in a binary system, or flare activity. Finally, although many late-M dwarfs in the Upper Scorpius ($\sim$5--10 Myr) association have strong H$\alpha$ emission, only a few of the early L-type objects of this region also present noticeable H$\alpha$ emission, such as VISTA J1607$-$2146, VISTA J1611$-$2215, and VISTA 1615$-$2229 \citep{Lodieu2018}.

In addition, some candidate young exoplanets still in their formation stage have been found to present strong H$\alpha$ emission, probably caused by accretion. This is the case of PDS 70 b and c \citep{Wagner2018, Haffert2019}, and LkCa15 \citep{Kraus2012, Sallum2015}, although recent studies showed that the H$\alpha$ emission in this last object may come from an inner disk structure and not from a planet \citep{Thalmann2016, Mendigutia2018, Currie2019}. Some techniques have been proposed for using this H$\alpha$ emission as a resource to detect new young exoplanets, such as spectro-astrometry \citep{Mendigutia2018}, or adaptive optics imaging in the H$\alpha$ band \citep[e.g. ][]{Close2014, Huelamo2018, Cugno2019, Zurlo2020}.

2MASS~J0249$-$0557~c is very similar in age, mass, and temperature to the planet $\beta$\,Pic\,b \citep{Lagrange2010}. This object may therefore show a similarly strong H$\alpha$ emission. On the other hand, another similar wide companion in the $\beta$ Pic moving group is 2MASS J21265040$-$8140293. This object has a spectral type of L3 and orbits the $\beta$ Pic candidate member TYC 9486-927-1 at a separation of $>4500$\,AU \citep{Deacon2016}.\footnote{{As pointed out in \citet{Deacon2016}, this system is also compatible with the Tucana-Horologium association. However, it most likely belongs to the $\beta$ Pic association.}} However, 2MASS J21265040$-$8140293 does not show any remarkable H$\alpha$ emission (pEW $> -15\AA$) in its optical spectrum \citep{Cruz2007}\footnote{{Spectrum available at the BDNYC database in \url{http://database.bdnyc.org/browse}}}. The planet $\beta$\,Pic\,b has only been spectroscopically observed in the near-infrared \citep{Chilcote2017}, and no optical spectrum has been obtained so far. The possibility of an H$\alpha$ emission in this planetary mass object therefore has not been tested yet.

In conclusion, strong H$\alpha$ emission such as is observed in 2MASS J0249$-$0557 c is uncommon among old field L dwarfs, but several cases of young L-type objects exhibiting H$\alpha$ emission at similar strength are known. However, in most of these examples, the objects are younger than 10\,Myr, and this  may be related to the presence of accretion disks. 
At the age of the $\beta$ Pic ($\sim$20--25 Myr), the gas in the disks is expected to have dissipated \citep{Haisch2001}. Nevertheless, this may not be the case for lower mass substellar objects, which may have longer disk-decay timescales \citep{Riaz2008}. Some cases of $\beta$ Pic brown dwarf members show signs of the presence of a disk and ongoing accretion \citep[e.g. 2M0335+23, 2M1935$-$28;][]{Shkolnik2009, Shkolnik2012, Liu2016}, therefore this possibility cannot be ruled out. The lack of IR excess in the \textit{WISE} $w1$ and $w2$ photometry of the object (see Table \ref{Table:gendata}) indicates the absence of a warm disk around it. However, although the origin of the H$\alpha$ emission may be chromospheric, we cannot rule out a colder disk of material or the possibility of gas accretion in the object. As an example, some of the planetary mass objects with signs of accretion in Sigma Orionis, such as S\,Ori\,55, only show infrared excesses at wavelengths longer than 5 microns \citep{Zapatero2007b}.

    \begin{figure*}
   \centering
   \includegraphics[width=13.5cm]{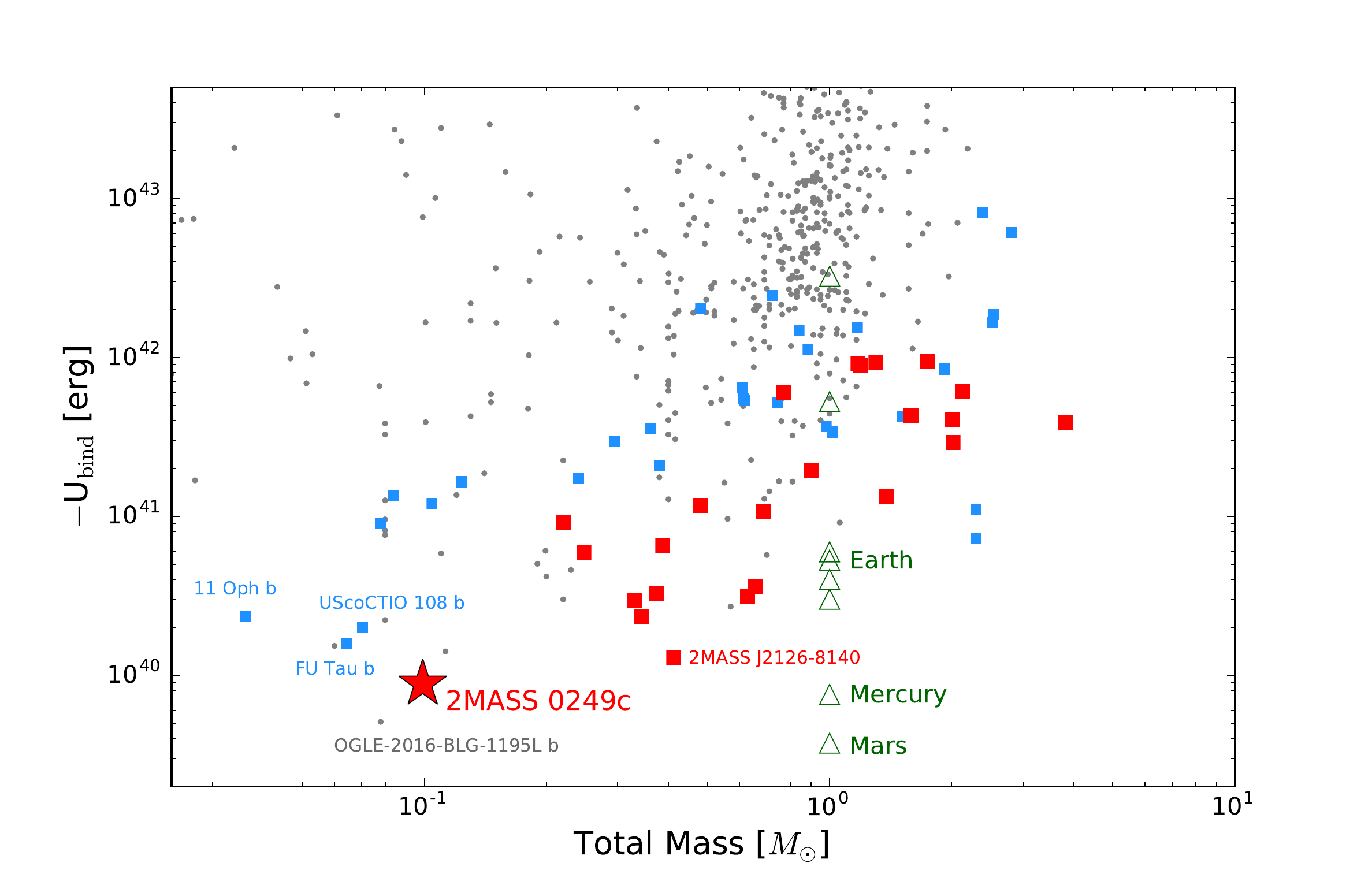}
   \caption[]{Binding energy compared to the total mass of the system for known substellar companions, including planets and brown dwarfs. The binding energy of 2MASS J0249$-$0557~c is marked with a red star. Other substellar companions with separations $>$1000 AU are marked as red squares (masses and separations from \citet{Chinchilla2020} and references therein). Objects with orbital separations $<$100 AU are marked as grey dots, and objects with separations between 100--1000 AU are marked as small blue squares (masses and separations from The Extrasolar Planets Encyclopaedia\footnotemark[4]; \citet{exoplaneteu}).  The binding energies of the Solar System planets are also included, marked as green triangles.}
              \label{Figure:binding}%
    \end{figure*}

\subsection{Binding energy and tidal disruption}

Widely separated systems, especially those with low binding energies, are prone to disruption by external perturbations, such as close encounters with other stars \citep{Kroupa1995, Close2007}. This is the case of the 2MASS J0249$-$0557 c component.

We calculated the binding energy of the system, as described in \citet{Chinchilla2020}. Masses and separations were obtained from \citet{Dupuy2018}. We infer an (absolute) upper limit for the binding energy of 2MASS J0249$-$0557 c of $U=(-8.8\pm4.4) 10^{32}$ J. This result considers an orbital semi-major axis equal to the current observed separation. However, the actual semi-major axis may be larger due to projection effects. When we consider an average semi-major axis of $\langle a_{rel} \rangle = 1.26\, d \, \langle \alpha \rangle$ \citep{Fischer1992}, where $d$ is the heliocentric distance and $\langle \alpha \rangle$ is the average observed angular separation of the components, the binding energy would become $U\sim(-7.0\pm3.4) 10^{32}$ J. This binding energy is surprisingly low and makes this system one of the most fragile wide systems discovered up to date. Figure \ref{Figure:binding} shows the binding energy of 2MASS~J0249$-$0557\,AB(c) in comparison with other very wide ($>$1000\,AU) binaries, exoplanets, and the planets of our Solar System. This figure shows that 2MASS~J0249$-$0557\,AB(c) is one of the substellar companions with the lowest binding energy known to date.

In order to estimate the time that this system is expected to survive in its actual environment,  we calculated the expected disruption time using the relations from \citet{BinneyTremaine} and \citet{Weinberg1987}, also as described in \citet{Chinchilla2020}. We considered a similar mean velocity dispersion as the solar neighbourhood, calculated using the data in \citet{Zapatero2007}; we obtain a value of 52.8 km s$^{-1}$. To obtain the density of stars around the system, we used $Gaia$ DR2 to select all the objects in a sphere of 8 pc radius around the system. We found 138 objects in this volume, which corresponds to a density of 0.064 obj pc$^{-3}$. Using these values, we obtain an estimated disruption time of 4--14 Gyr. We must consider that this system may have formed in a denser environment, similar to young forming regions and associations. Considering a higher density of objects ($\sim$10 obj pc$^{-3}$), we obtain a disruption time of 30--60\,Myr for this system. These two results combined indicate that the system may have been formed in the outskirts of a young open cluster or stellar association, being physically bound since then, and that it will probably survive to ages similar to that of our Solar System. It is interesting to note that 2MASS~J0249$-$0557\,c, with an estimated mass of $\sim$12\,M$_{\mathrm{Jup}}$, will have a temperature of $\sim$\,500\,K at the age of the Sun, within the Y dwarf spectral type range, which makes the detection of similar systems at these old ages very difficult with current means. The James Webb Space Telescope (JWST), with its near. and mid-infrared capabilities, may be able to identify these types of objects.


\section{Summary and final remarks}

We presented the independent identification of 2MASS~J0249$-$0557\,c: a wide (2640$\pm$20\,AU) substellar companion of the binary brown dwarfs 2MASS~J0249$-$0557\,AB, recently discovered by \citet{Dupuy2018}. We found this candidate companion in our proper motion search around YMG members by combining the astrometry and photometry of the VHS and 2MASS catalogues, and using photometry from other catalogues such as DENIS and/or AllWise, and astrometry from $Gaia$ DR2, when available.

\footnotetext[4]{\url{http://exoplanet.eu}}

  We acquired optical and near-infrared spectroscopy of 2MASS~J0249$-$0557\,c to characterise its spectral energy distribution and atmospheric properties. We assigned a spectral type of L2.5$\pm$0.5 in the optical and L3$\pm$1 in the near-infrared. This spectral classification is consistent with that of \citet{Dupuy2018}.
  
    We found spectral features characteristic of youth, such as strong H$\alpha$ emission in the optical, a peaked shape in the $H$ band, weak alkaline lines, and strong VO, TiO, and FeH bands both in the optical and near-infrared. We observe a tentative Li I detection with a pEW of $\lesssim$ 5~$\AA$. These features are consistent  with the very low gravity classification by \citet{Dupuy2018} and with the expected membership in $\beta$ Pic.
    
      2MASS~J0249$-$0557\,c shows a strong H$\alpha$ emission line, with a pEW of $-$90$^{+20}_{-40}$, which is likely stable on timescales of hours and years. This emission is not very common among old field L dwarfs, but it may be present in some of the young early-L dwarfs, especially those located in regions of ages younger than 10\,Myr. This reinforces the young age of 2MASS~J0249$-$0557\,c.
      
    With an (absolute) upper limit of $U=(-8.8\pm4.4) 10^{32}$J, 2MASS~J0249$-$0557\,AB(c) is one of the widest low-mass systems with the lowest binding energy known to date. It may have been formed in the outskirts of a young star formation region and it is expected to survive at ages similar to our Solar System. The solar vicinity may be populated by similar systems, but they may have remained unnoticed with current facilities.    

\begin{acknowledgements}
     We thank the anonymous referee for his/her very useful corrections and suggestions, which helped improving this manuscript. This paper is based on observations performed at the European Southern Observatory (ESO) in Chile, under programme 0101.C-0389, PI: P. Chinchilla. This work is based on observations (program GTC06-18BDDT, PI: P. Chinchilla) made with the Gran Telescopio Canarias (GTC), operated on the island of La Palma in the Spanish Observatorio del Roque de los Muchachos of the Instituto de Astrof\'isica de Canarias. The authors want to thank Dr. Antonio Cabrera for performing the DDT GTC/OSIRIS observations, and Prof. Rafael Rebolo and Prof. Eduardo Mart\'in for their useful comments and discussions. P.C., V.J.S.B. and N.L. were financially supported by the program PID2019-109522GB-C53, and M.R.Z.O. by the program PID2019-109522GB-C51 of the Ministerio de Ciencia e Innovaci\'on. 
This research has made use of the Simbad and Vizier databases, operated
at the centre de Donn\'ees Astronomiques de Strasbourg (CDS), and
of NASA's Astrophysics Data System Bibliographic Services (ADS).
This research has also made use of some of the tools developed as part of the
Virtual Observatory.
This work has made use of data from the European Space Agency (ESA) mission
{\it Gaia} (\url{https://www.cosmos.esa.int/gaia}), processed by the {\it Gaia}
Data Processing and Analysis Consortium (DPAC,
\url{https://www.cosmos.esa.int/web/gaia/dpac/consortium}). Funding for the DPAC
has been provided by national institutions, in particular the institutions
participating in the {\it Gaia} Multilateral Agreement.
This publication makes use of data products from the Two Micron All Sky Survey, which is a joint 
project of the University of Massachusetts and the Infrared Processing and Analysis 
Center/California Institute of Technology, funded by the National Aeronautics and Space 
Administration and the National Science Foundation. 
This publication has made use of the Young Brown Dwarf Compilation maintained by the BDNYC collaboration led by Kelle Cruz, Emily Rice, and Jackie Faherty. This publication made use of Python programming language (Python Software Foundation, \url{https://www.python.org}).

\end{acknowledgements}







\bibliographystyle{aa} 
\bibliography{biblio.bib} 

\end{document}